\documentclass[reprint,superscriptaddress,twocolumn]{revtex4-1}
\usepackage{amsmath,amssymb,mathrsfs}
\usepackage{braket}
\usepackage{hyperref}
\usepackage{paralist}
\usepackage{bm}
\usepackage{graphicx}
\usepackage{setspace}
\usepackage{xcolor,soul}
\sethlcolor{cyan}

\begin{document}

%%=================================================================================

\title{Cooling and entangling ultracold atoms in optical lattices}

%%=================================================================================

\author{Bing Yang}
\affiliation{Hefei National Laboratory for Physical Sciences at Microscale and Department of Modern Physics, University of Science and Technology of China, Hefei, Anhui 230026, China}
\affiliation{Physikalisches Institut, Ruprecht-Karls-Universit\"{a}t Heidelberg, Im Neuenheimer Feld 226, 69120 Heidelberg, Germany}
\affiliation{CAS Centre for Excellence and Synergetic Innovation Centre in Quantum Information and Quantum Physics, University of Science and Technology of China, Hefei, Anhui 230026, China}
\author{Hui Sun}
\affiliation{Hefei National Laboratory for Physical Sciences at Microscale and Department of Modern Physics, University of Science and Technology of China, Hefei, Anhui 230026, China}
\affiliation{Physikalisches Institut, Ruprecht-Karls-Universit\"{a}t Heidelberg, Im Neuenheimer Feld 226, 69120 Heidelberg, Germany}
\affiliation{CAS Centre for Excellence and Synergetic Innovation Centre in Quantum Information and Quantum Physics, University of Science and Technology of China, Hefei, Anhui 230026, China}
\author{Chun-Jiong Huang}
\affiliation{Hefei National Laboratory for Physical Sciences at Microscale and Department of Modern Physics, University of Science and Technology of China, Hefei, Anhui 230026, China}
\affiliation{CAS Centre for Excellence and Synergetic Innovation Centre in Quantum Information and Quantum Physics, University of Science and Technology of China, Hefei, Anhui 230026, China}
\author{Han-Yi Wang}
\affiliation{Hefei National Laboratory for Physical Sciences at Microscale and Department of Modern Physics, University of Science and Technology of China, Hefei, Anhui 230026, China}
\affiliation{Physikalisches Institut, Ruprecht-Karls-Universit\"{a}t Heidelberg, Im Neuenheimer Feld 226, 69120 Heidelberg, Germany}
\affiliation{CAS Centre for Excellence and Synergetic Innovation Centre in Quantum Information and Quantum Physics, University of Science and Technology of China, Hefei, Anhui 230026, China}
\author{Youjin Deng}
\affiliation{Hefei National Laboratory for Physical Sciences at Microscale and Department of Modern Physics, University of Science and Technology of China, Hefei, Anhui 230026, China}
\affiliation{CAS Centre for Excellence and Synergetic Innovation Centre in Quantum Information and Quantum Physics, University of Science and Technology of China, Hefei, Anhui 230026, China}
\author{Han-Ning Dai}
\affiliation{Hefei National Laboratory for Physical Sciences at Microscale and Department of Modern Physics, University of Science and Technology of China, Hefei, Anhui 230026, China}
\affiliation{Physikalisches Institut, Ruprecht-Karls-Universit\"{a}t Heidelberg, Im Neuenheimer Feld 226, 69120 Heidelberg, Germany}
\affiliation{CAS Centre for Excellence and Synergetic Innovation Centre in Quantum Information and Quantum Physics, University of Science and Technology of China, Hefei, Anhui 230026, China}
\author{Zhen-Sheng Yuan}
\email[e-mail:]{yuanzs@ustc.edu.cn}
\affiliation{Hefei National Laboratory for Physical Sciences at Microscale and Department of Modern Physics, University of Science and Technology of China, Hefei, Anhui 230026, China}
\affiliation{Physikalisches Institut, Ruprecht-Karls-Universit\"{a}t Heidelberg, Im Neuenheimer Feld 226, 69120 Heidelberg, Germany}
\affiliation{CAS Centre for Excellence and Synergetic Innovation Centre in Quantum Information and Quantum Physics, University of Science and Technology of China, Hefei, Anhui 230026, China}
\author{Jian-Wei Pan}
\email[e-mail:]{pan@ustc.edu.cn}
\affiliation{Hefei National Laboratory for Physical Sciences at Microscale and Department of Modern Physics, University of Science and Technology of China, Hefei, Anhui 230026, China}
\affiliation{Physikalisches Institut, Ruprecht-Karls-Universit\"{a}t Heidelberg, Im Neuenheimer Feld 226, 69120 Heidelberg, Germany}
\affiliation{CAS Centre for Excellence and Synergetic Innovation Centre in Quantum Information and Quantum Physics, University of Science and Technology of China, Hefei, Anhui 230026, China}

%\maketitle

%%=================================================================================
\begin{abstract}
Scalable, coherent many-body systems are of fundamental interest, which enable the realization of new quantum phases and could exponentially speed up the information processing. Here we report the cooling of a quantum simulator with ten-thousand atoms and mass production of high-fidelity entangled pairs.
In a two-dimensional plane, we cool the Mott-insulator samples by immersing them into removable superfluid reservoirs and a record-low entropy per particle of  $1.9^{+1.7}_{-0.4} \times 10^{-3} k_{\text{B}}$ is achieved.
The atoms are then rearranged into a two-dimensional lattice free of defects.
We further demonstrate a two-qubit gate with a fidelity of $0.993(1)$ for entangling 1250 atom pairs.
This experiment offers a way for exploring low-energy many-body phases and could enable the creation of a large-scale entanglement.
\end{abstract}
%%=================================================================================

\maketitle

Cooling many-body systems to ultracold regimes has revolutionized the field of quantum physics.
A paradigmatic quantum simulator is composed of ultracold neutral atoms in optical lattices, which provide a clean and controllable platform for studying complex many-body problems \cite{Bloch:2008,Greiner:2002}.
However, the thermal entropy arising from the intrinsic heating \cite{Pichler:2010,Trotzky:2010}, non-adiabaticity \cite{Chiu:2018,Dolfi:2015} and inefficient thermalization \cite{Hung:2010} hinders the revealing of novel phases of matter \cite{McKay:2011} and scaling up the entangled states \cite{Nielsen:2010,Shor:1996}.
Besides numerous efforts that have been applied to cool the quantum gases \cite{Catani:2009,Tin-LunHo:2009,Bakr:2010,Medley:2011,Taie:2012,Greif:2013}, some theoretical schemes suggest to immerse lattice trapped atoms into superfluid reservoirs which could eventually carry away the thermal entropy \cite{Griessner:2006,Kantian:2018}.
For example, immersing the atomic sample in a removable surrounding reservoir has enabled the observation of a Fermi-Hubbard antiferromagnet with a quantum gas microscope \cite{Mazurenko:2017,Chiu:2018}.
However, either mixing two distinct quantum phases for efficient thermalization or realizing sub-lattice addressability to remove the reservoir, poses outstanding experimental challenges.

%%=================================================================================

\begin{figure}[!htb]
\centering
 {\includegraphics[width=7.26cm]{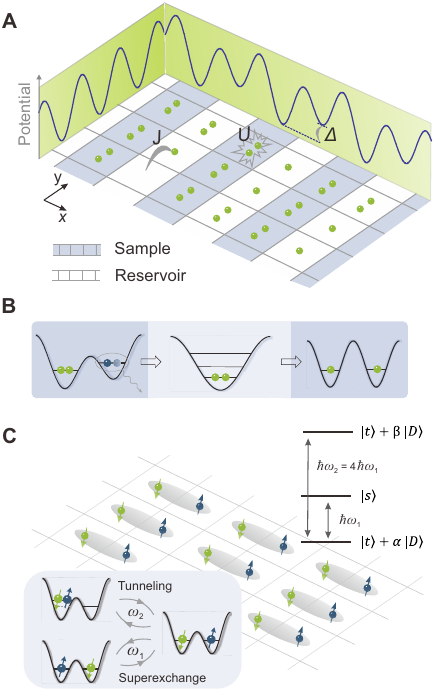}}
  \caption{Cooling and entangling atoms. (A) Schematic of the staggered-immersion cooling. In an optical superlattice, The gapped Mott insulators are immersed in the gapless superfluid reservoirs with a staggered geometry.
  (B) Redistribution of the atoms into a unity filling state by engineering the lattice potential.
  (C) Entangling thousands of atom pairs in parallel. When the initial state in double-wells (DWs) is $\ket{\downarrow,\uparrow}$, the Hilbert space contains three eigenstates. The triplet/singlet states are $\ket{t/s}= (\ket{\downarrow,\uparrow} \pm \ket{\uparrow,\downarrow})/\sqrt{2}$, the double-occupied  state is $\ket{D}=(\ket{\uparrow\downarrow,0} + \ket{0,\uparrow\downarrow})/\sqrt{2}$, and the coefficients $\alpha$ and $\beta$ depend on $J$ and $U$. At $J/U=\sqrt{3}/4$, the frequencies satisfy a relation of $\omega_2 = 4 \omega_1$. A two-qubit $\sqrt{\text{SWAP}}$ gate is realized after the evolution time of $\pi/(2\omega_1)$. }
\label{Fig:f1}
\end{figure}

%%=================================================================================

Here we design a staggered lattice to immerse the samples alternately into superfluid reservoirs for sufficient thermalization.
Meanwhile, employing a sub-lattice addressing technique, all the high-entropy reservoirs are removed and the low-entropy regime is achieved.
The cooling concept is sketched in Fig. \ref{Fig:f1}A, where the superfluid that holds large density of states serves as a reservoir for storing the entropy of the joint system.
In our 2D bosonic system with a bichromatic superlattice along the $x$ direction, the competition between the kinetic energy $J$ and interaction energy $U$ leads to the formation of superfluid and Mott insulator phases.
An alternating appearance of the insulator and superfluid is realized by adjusting the local chemical potentials.
Based on the local density variations, we develop a thermometry method to characterize the cooling performance.
After isolating and removing the particles in the reservoirs, the atoms of the samples (Fig. \ref{Fig:f1}B) are redistributed within the superlattice to achieve a unity-filling state with over $10^4$ sites.

The defect-free platform after cooling is a great starting point for quantum information processing.
Owing to the parallel controllability of atoms in optical lattices, multipartite entanglement can be created by coupling the atomic qubits with nearest-neighbor interactions \cite{Jaksch:1999,Duan:2003}.
Although the single- and two-qubit control \cite{Weitenberg:2011,Wang:2016,Anderlini:2007,Trotzky:2008,Dai:2016} have been demonstrated in this system, the current challenge towards large-scale entanglement is to develop a high-fidelity entangling gate \cite{Ladd:2010,Raussendorf:2001}.
Here we enhance the superexchange interaction by engineering the Hamiltonian in a three-level scheme (Fig. \ref{Fig:f1}C).
The dynamical evolution is driven by the coherent competition of superexchange and atom tunnelling.
Compared with the earlier experiment \cite{Dai:2016}, we dramatically reduce the gate operation time and thereby significantly improve the operation fidelity, exceeding the threshold of some fault-tolerant models for quantum computing \cite{Fowler:2012,Knill:2005,Raussendorf:2007}.

%%=================================================================================

\begin{figure*}[!htb]
{\includegraphics[width=12.1cm]{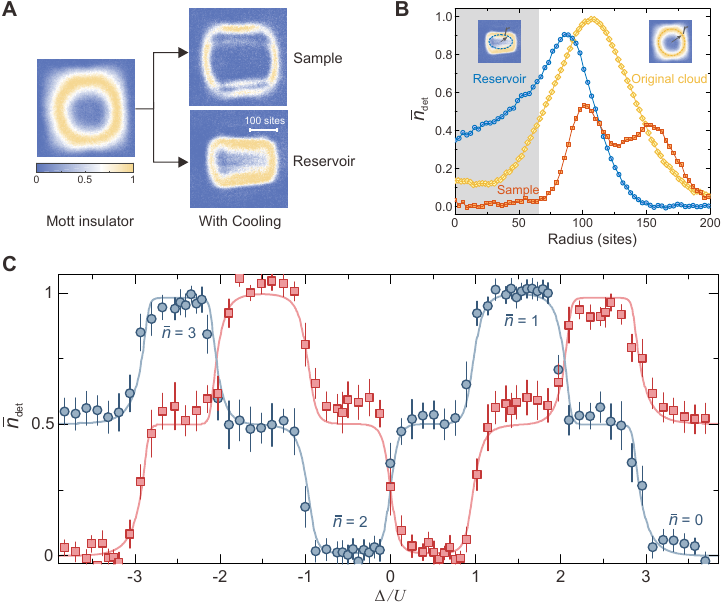}}
  \caption{Cooling a quantum gas.
  (A) Exemplary images of the atomic densities averaged over 50 measurements. The original Mott insulator is created in the short-lattices without cooling. After applying the cooling sequence, we detect the odd (sample) and even (reservoir) subsystems separately. With a parity projection for the $\bar{n}=2$ Mott insulator, less remaining density in the cloud center indicates lower entropy.
  (B) The density profiles are azimuthal averages over the corresponding images in (A), as sketched in the insets. For calculating the mean densities in the samples (red) and the reservoirs (blue), radius $r$ refers to the geometric mean of the elliptical radii (the aspect ratio for this ellipse is 0.6).The entropy reduction is significant in the shaded area containing $10^4$ lattice sites.
  (C) Density partitions between the central region of the odd (blue) and even (red) subsystems.
  The superfluid and Mott insulator phases are coexistent and emerge alternately, which leads to the changes of density from 3.5 (0) to 0 (3.5). The staircase-like changing of the densities is a signature of the low temperature. The solid curves are isentropic simulations at $k_B T_f/U = 0.05$. The subtle features, such as larger statistic errors of the superfluid states, higher temperatures at larger $|\Delta|$ and unequal width of the Mott plateaus are revealed by our data and simulations. Each of the error bars represents the standard deviation over 20 images.}
\label{Fig:f2}
\end{figure*}

%%=================================================================================

The experiment starts with a $^{87}$Rb Bose-Einstein condensate of $\sim$8.6$\times 10^4$ atoms confined in a single well of a pancake-shaped standing wave \cite{Dai:2016}.
To implement the cooling procedure, we ramp up the lattices to create a phase separation between the subsystems.
The 2D cloud is adiabatically loaded into a square lattice in the $x$-$y$ plane with a period of $\lambda_s/2$, here $\lambda_s = $ 767 nm denotes the wavelength of the ``short'' lattices.
Another ``long'' lattice with wavelength $\lambda_l=2\lambda_s$ is employed to construct the superlattice that separates the quantum gas into odd and even subsystems.
Their lattice depths are ramped exponentially in 60 ms with a time constant of 20 ms up to $V_s = 26.1(2) E_r$ and $V_l = 10.0 (1)\ E_r$, where $E_r = h^2/(2 m \lambda_s^2)$ is the recoil energy with $h$ the Planck constant and $m$ the atomic mass.
Meanwhile, the transverse trap is adjusted to set the atomic densities to $\bar{n} \simeq 1.75$ in the central area.
The energy offset $\Delta$ is controlled by shifting the relative phase of the superlattice lasers \cite{Dai:2016}.
As the tunneling strength decreases across the critical value, gapped Mott insulators with $\bar{n} =2$ start to emerge from the contacting superfluid states with $\bar{n} \simeq 1.5$.

After the atoms enter the deep lattices, coherent tunneling is negligible and defects in Mott insulators are induced only by thermal fluctuations \cite{Bakr:2010}.
The on-site number fluctuations is probed via the photoassociation collisions, with which the occupation of a site is reduced to its odd-even parity.
The odd and even subsystems are distinguished by turning on a spin-dependent superlattice to separate their hyperfine transition by 28 kHz \cite{Dai:2016}.
Then, we perform a Landau-Zener sweep to transfer the atoms of the even rows from the initial state $\ket{\downarrow} \equiv \Ket{F=1, m_F=-1}$ to $\ket{\uparrow} \equiv \Ket{F=2, m_F=-2}$, with an efficiency of 99.5(3)\% \cite{SM}.
The \emph{in situ} atomic densities $\bar{n}_{det}$ of the odd and even subsystems (Fig. \ref{Fig:f2}A) are successively recorded by absorption imaging with an optical resolution of $\sim$1.0 $\mu$m.
Fig. \ref{Fig:f2}B shows the density profiles, where the number fluctuations of the samples decrease dramatically compared to the Mott insulator without applying the staggered-immersion cooling.
Nevertheless, the efficiency of the parity projection 98.7(3)\% leads to an undesired residual to the measured density, which constrains the lowest deduced entropy from such remaining density to $0.011(2) k_{\text{B}}$.
To overcome this imperfection for determining the ultra-low temperature $T_f$ of the target samples ($\Delta = -U/2$), we apply thermometry on the samples via probing the densities in the subsystems.

The partitions of atomic densities are measured by scanning the offset of the local chemical potential.
Fig. \ref{Fig:f2}C shows the densities at different $\Delta/U$, where the Mott insulators emerge associated with superfluid states in the other subsystem.
Changing $\Delta$ represents a small perturbation to the lattice, thus the size of the whole cloud is maintained and only local transport is allowed.
Eight plateaus emerge around the half-integers of $\Delta/U$, which are analogous to the concentric shell structures of the Mott insulator in a harmonic trap \cite{Bakr:2010}.
Intriguingly, the compressible superfluid states exhibit distinctive half-integer plateaus owing to the local particle conservation.
In the absence of global mass flow, the entropy is most likely to be conserved locally and meanwhile the nonadiabaticities of the lattice loading are minimized \cite{Dolfi:2015}.
In addition, we measure the thermalization rate by monitoring a quench dynamics (reverse the sign of $\Delta$ at certain times during the lattice ramping), and find that the system can equilibrate on the time scale of $h/J$ \cite{SM}.

%%=================================================================================

\begin{figure}[!htb]
\centering
{\includegraphics[width=6.15cm]{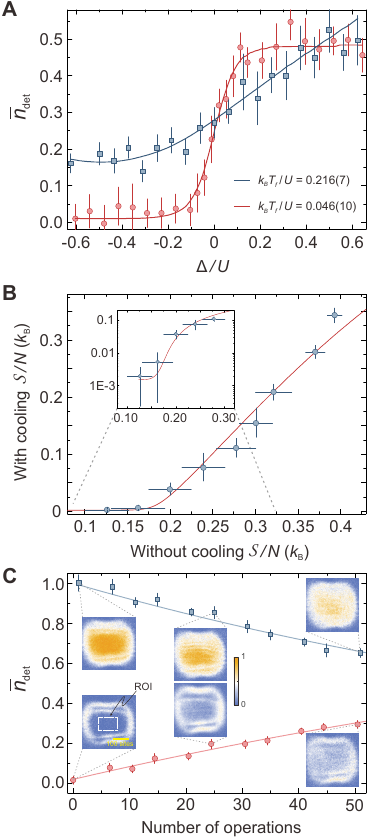}}
  \caption{Thermometry of the quantum gases and atom redistribution.
  (A) Density variations at 60 ms and 500 ms lattice ramping. The corresponding curves are isentropic fitting results.
  (B) Cooling efficiency and the critical behavior. The red curve is the theoretical prediction based on Monte-Carlo simulations. The inset shows the entropy of the system with cooling in logarithmic scale.
  (C) Calibrating the fidelity of atom redistribution. We repeat the operation from $|2,0\rangle$ to $|1,1\rangle$ and its reversal in cycles, then count the atoms on $10^4$ lattice sites within a region of interest (ROI). Each data point is averaged over 20 measurements and the error bars denote the standard deviations. The red and blue curves are power-law fitting results. }
\label{Fig:f3}
\end{figure}

%%=================================================================================

Temperature of the samples is fitted based on particle and entropy conservation at different $\Delta/U$.
Fig. \ref{Fig:f3}A shows $\bar{n}_{\text{det}}$ of the odd subsystem as a function of $\Delta/U$ at two experimental settings.
The phase transition at the edge of the plateaus leads to sudden changes of the occupancies, which allows the thermometer to access ultra-low temperatures.
In these samples, the lowest temperature achieved is $k_\text{B} T_f=0.046(10)U$.
To characterize the cooling performance, we control the entropy by changing the adiabaticity or duration of the lattice loading \cite{Pichler:2010,Dolfi:2015}.
For each superlattice loading sequence, a Mott insulator with central density $\bar{n}=2$ is prepared by setting $\Delta$=0 to disable the cooling effect.
Thereby the entropy in the absence of cooling is deduced from the parity measurement.
As shown in Fig. \ref{Fig:f3}B, only if the entropy without cooling achieves a critical value $\mathcal{S}/N \sim 0.17 k_{\text{B}}$, the cooling power becomes strong and the entropy of the samples reduces significantly.
This corresponds to a phase transition of the reservoir from a normal fluid to a superfluid \cite{Trotzky:2010}, which shifts to a value below $k_{\text{B}} \ln2/(2\bar{n}) = 0.20 k_{\text{B}}$ as the system equilibrates at non-zero tunneling strength \cite{SM}.
The experimental results are in excellent agreement with the quantum Monte-Carlo simulations.
In addition, the temperatures of smaller samples can be determined from the density profiles, which are consistent with the thermometry method above.
Losing less than half of the total atoms in the reservoir ($\sim$1.5/3.5), we lower the entropy per particle in the large samples to $\mathcal{S}/N = 1.9^{+1.7}_{-0.4} \times 10^{-3} k_{\text{B}}$, corresponding to a 65-fold reduction of the original entropy.
Besides the entropy inherent in the 2D gas, the intrinsic heating induced by the light-induced scattering during the lattice loading process is $0.025(2) k_{\text{B}}$ per particle \cite{SM}, indicating that the cooling power has exceeded the heating rate of the optical lattices.

%%=================================================================================

\begin{figure*}[!htb]
{\includegraphics[width=14.6cm]{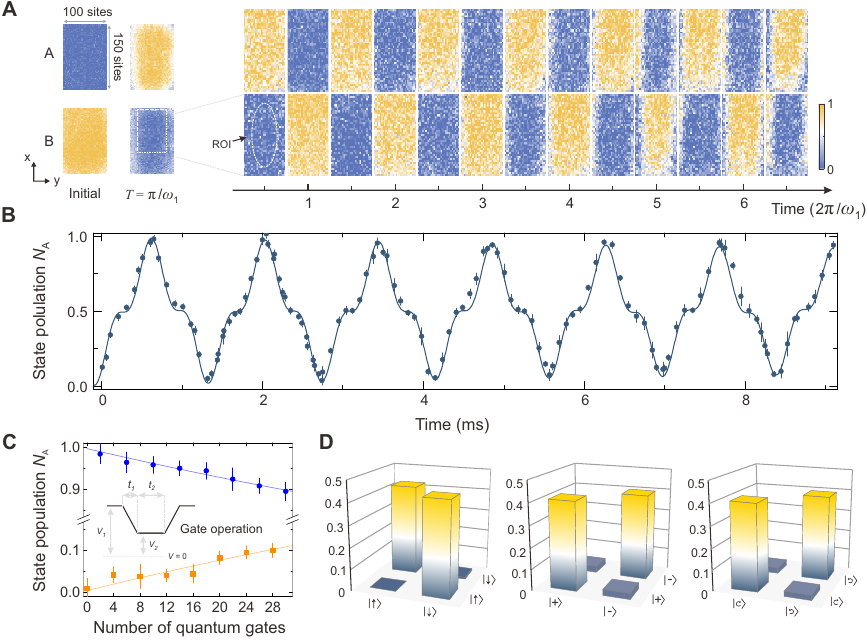}}
  \caption{Two-qubit quantum gate.(A) Spin dynamics. We record the observables $N_A = \langle n_{\uparrow,\text{odd}} + n_{\downarrow,\text{even}}\rangle/2$ on image A and $N_B = \langle n_{\downarrow,\text{odd}} + n_{\uparrow,\text{even}}\rangle/2$ on image B, successively.
  The exemplary images show the initial state and the one after half superexchange period. The elliptical region of interest (ROI) containing 2500 lattice sites is chosen for counting signals. The high contrast of each groups of the images indicates excellent coherence of the system. The spatial inhomogeneity of the interactions among the sample starts to emerge after multi-periods.
  (B) Time-resolved observation of spin dynamics. The state evolution shows a coherent interference between two processes. Every data point is averaged over 5 repetitions among the ROI. The solid curve is a fitting result by considering the coherent evolution and a phenomenological exponential decay, which gives $\omega_1/2\pi = 709(3)$ Hz and the 57(16) ms lifetime.
  (C) Gate operation and state coherence. The quantum gate is implemented by controlling the depth of the short-lattice, where $V_1$ = 31.2(2) $E_r$,  $V_2$ = 9.72(7) $E_r$, $t_1$ = 250 $\mu s$ and $t_2$ = 270 $\mu s$. We apply multiple two-qubit gates and then record the state populations. Each data point is averaged over 20 measurements and the error bars are the standard deviations. The blue and orange curves are power-law fitting results.
  (D) Spin correlations. The spin state after 29 times of gate operations is measured on $|\downarrow\rangle/|\uparrow\rangle$, $|+\rangle/|-\rangle$ and $|\circlearrowright\rangle/|\circlearrowleft\rangle$ basis.}
\label{Fig:f4}
\end{figure*}

%%=================================================================================

We implement a state engineering over $10^4$ sites to turn the low-entropy samples into a unity-filling state.
This is achieved by parallel manipulating the DWs after cleaning the particles in the high-entropy reservoirs.
Fig. \ref{Fig:f1}B shows the sequences from the number state $|2,0\rangle$ to $|1,1\rangle$ in an isolated DW unit.
To avoid level crossing causing band excitation, we shift the centroid of atom pairs to match the antinodes of the long-lattice before transferring the atoms adiabatically onto its ground band.
Next, we ramp up the short-lattice to split the atom pairs into a product state with one atom per site.
The atoms favor to occupy different sites in the presence of the interaction blockade $U$.
The achieved fidelity of 99.3(1)\% is calibrated by applying multiple operations to the samples, i.e. driving the atoms between $|2,0\rangle$ and $|1,1\rangle$ back and forth as in Fig. \ref{Fig:f3}C.
Together with the entropy from the samples, the overall fidelity for preparing a unity filling state over $10^4$ lattice sites is 99.2(1)\%.

%%=================================================================================

The defect-free platform provides massive physical qubits for quantum information processing \cite{Ladd:2010,Raussendorf:2001,Dai:2016}.
We implement the entangling gate within all the DWs using the three-level scheme.
For probing the spin dynamics, the state initialization and detection have the same procedure as our previous experiment \cite{Dai:2016}.
Here, the potentials are set to 9.76(1) $E_r$ for short-lattice and 10.0(1) $E_r$ for long-lattice.
In Fig. \ref{Fig:f4}B, we observe an interference between two processes for 6.5 superexchange periods, which is equivalent to the application of 26 gate operations.
The lifetime 57(16)ms of the spin dynamics is mainly caused by the spatial inhomogeneity of the interactions and fluctuations of the superlattice phase \cite{Dai:2016}.
Besides these technical issues, we further study the decoherence caused by light scattering and achieve a coherent oscillation with a lifetime of 646(32) ms.

The gate fidelity is characterized by measuring the target Bell state $\ket{t}$, which is acquired by performing a two-qubit gate followed by a phase rotation \cite{SM}.
The fidelity of the experimentally produced density matrix is $F =  \bra{t}\rho^{\text{exp}}\ket{t} = \frac{1}{2}(P_{\ket{\uparrow,\downarrow}} +P_{\ket{\downarrow,\uparrow}}) + \frac{1}{4}(\langle\sigma_x \sigma_x\rangle + \langle\sigma_y \sigma_y\rangle)$, where $\sigma_{x,y}$ are Pauli spin operators.
The bases for the $\sigma_x$ and $\sigma_y$ operators are $\ket{+/-}=(\ket{\uparrow}\pm\ket{\downarrow})/\sqrt{2}$ and $\ket{\circlearrowright/\circlearrowleft}=(\ket{\uparrow}\pm i\ket{\downarrow})/\sqrt{2}$, respectively.
The incoherent error of the gate operator accumulates and the fidelity scales as $F^N$ when multiple gates $N$ are applied.
From the measurements of the state populations (Fig. \ref{Fig:f4}C), we observe a reduction of the state coherence after implementing multiple quantum gates.
To avoid the influence of the detection error, we measure the fidelity of the Bell state after 29 consecutive gate operations and then deduce the single-gate fidelity.
Under the $\ket{\uparrow/\downarrow}$ basis, the probabilities of the states are $P_{\ket{\uparrow,\downarrow}}$ = 0.44(2), $P_{\ket{\downarrow,\uparrow}}$ = 0.42(1), $P_{\ket{\uparrow,\uparrow}}$ = 0.01(2), $P_{\ket{\downarrow,\downarrow}}$ = 0.01(3).
The states $\{\ket{\uparrow\downarrow,0}, \ket{0,\uparrow\downarrow}\}$ resulted from the operation error are included in the normalization.
For the basis of $\ket{+/-}$ and $\ket{\circlearrowright/\circlearrowleft}$, a microwave $\pi/2$ pulse is applied to the atoms and then the populations of the spin states are measured.
The $\sigma_y$ operator given by the pulse has a relative $\pi/2$ phase shift with the pulse of the $\sigma_x$ operator.
The expectation values of the spin correlations are $\langle\sigma_x \sigma_x \rangle$  = 0.75(3) and $\langle\sigma_y \sigma_y \rangle $ = 0.75(4).
The fidelity of this state after 29 gates is $0.80(2)$, from which the fidelity of an individual two-qubit gate is deduced as $F = 99.3(1)\%$.
The infidelity of this two-qubit gate is below the error threshold required by some fault-tolerant models \cite{Knill:2005,Raussendorf:2007,Fowler:2012}.

%%=================================================================================

In conclusion, we have demonstrated a cooling method to dramatically reduce the entropy of a quantum simulator.
The achieved entropy represents a new regime for studying the low-energy physics.
Furthermore, we have realized a high-fidelity two-qubit gate and prepared 1250 Bell states of neutral atoms in parallel.
The defect-free atom arrays and entangling gate are essential building blocks for creating multi-particle entanglement towards measurement-based quantum computing \cite{Raussendorf:2001,Raussendorf:2003,Vaucher:2008}.
For building a practical logic qubit towards fault-tolerant quantum computation \cite{Knill:2005,Raussendorf:2007,Fowler:2012}, our system meets the requirements on both the entangling gate and the number of physical qubits.
Our technique is generic and practical for cooling insulating states of other species and systems in other geometries \cite{Kantian:2018}.
For the gapless fermionic phases, our method could significantly improve the efficiency of entropy transport \cite{Tin-LunHo:2009,Mazurenko:2017}, giving access to lower-temperature regimes for observing the $d$-wave superfluid phase \cite{Hofstetter:2002}.
The defect-free platform and entangled pairs are suitable for exploring the bosonic antiferromagnetic order \cite{Soerensen:2010} and resonating valence bond states \cite{Trebst:2006}.

%%=================================================================================

%%%=================================================================================

%%%%%% Methods %%%%

\newpage
\onecolumngrid
\vspace*{0.5cm}
\begin{center}
    \textbf{SUPPLEMENTARY MATERIALS}
\end{center}
\vspace*{0.5cm}
%\twocolumngrid

\subsection{State preparation}
The 2D gas is prepared by adiabatically loading a nearly pure Bose-Einstein condensate on the $5S_{1/2}\Ket{F=1, m_F=-1}$ hyperfine state into the single layer of a pancake-shaped trap, which is a $3.0 \ \mu$m period standing-wave generated by interfering two blue-detuned laser beams at wavelength $\lambda_s$.
Then we levitate the atoms by ramping a magnetic gradient in 2.1 seconds up to 30.5 Gauss/cm to compensate the gravity.
The depth of the pancake trap is maintained at $2.3\ E_{r}$ during the levitation.
The transverse confinement on the $x-y$ plane is provided by a red-detuned 1070 nm laser with a Gaussian beam waist of 116 $\mu$m.
The trap frequencies of the 2D system are typically $\omega_{x,y,z} = 2\pi \times [7.8(3),7.8(3),1508(10)]$ Hz.
To reduce the lattice heating caused by the movement of the center of mass, the dipole oscillation of the ensemble in this 2D hybrid trap is minimized.

Next, we perform the lattice loading by simultaneously ramping up the trap potentials in 60 ms.
The pancake trap is ramped exponentially up to $201(1)\ E_{r}$ with a time constant of 20 ms.
The blue-detuned lattices have anti-trapping effects, while the red-detuned Gaussian beam provides a dipole trap for adjusting the central density.
In Fig.\ref{Fig:f1}C, the non-circular shape of the clouds results from these hybrid potentials.
The superlattice potential can be expressed as,
\begin{equation}
V(x)=V_s\mathrm{cos}^2(kx)-V_l\mathrm{cos} ^2(kx/2+\varphi).
\end{equation}
Here $k = 2\pi/\lambda_s$ is the wave number.
The relative phase $\varphi$ determines the superlattice structure, and $\varphi=0$ indicates a balanced condition for the double-well units.
We keep the phase $\varphi$ constant during the ramping process, and the energy shift at the end of the dynamics is represented by the offset $\Delta$.
After the lattice loading, the density distribution is frozen by increasing the short-lattices to $42\ E_r$.
We detect the $in\ situ$ densities via an absorption imaging with a microscope objective (N.A.=0.48) along the $z$ direction.

%%=================================================================================

\begin{figure*}[!htb]
{\includegraphics[width=15.5cm]{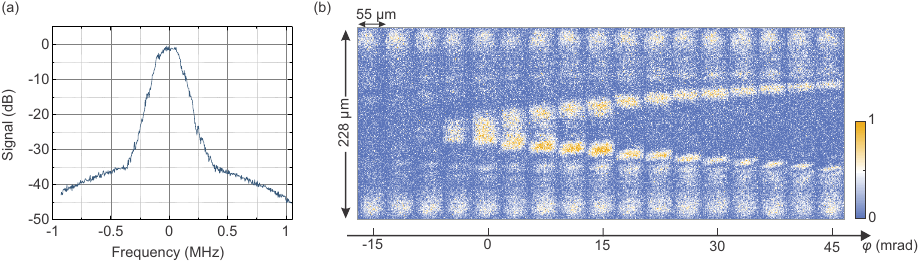}}
  \caption{Calibration of the superlattice phase. (a) Beat signal of the superlattice lasers. (b) Spatial inhomogeneity of the superlattice phase. The assembled image contains measurements at 17 different superlattice phases. Each subgraph has a size of $228 \times 55\ \mu m^2$ at the atomic plane. The appearance of a bunch of atoms at the upper or lower boundaries is caused by the shallow lattice potential. The high atomic densities around the central area indicates the zero point of the local superlattice phase.}
\label{Method:SLphase}
\end{figure*}

%%=================================================================================

%%=================================================================================

\subsection{Manipulation of atoms in the superlattice}

The superlattice phase $\varphi$ is controlled by changing the relative frequency of these lasers.
The 767 nm light is provided by a titanium-sapphire laser, whose frequency is stabilized by a reference cavity.
We lock the cavity onto the Rb D1 transition line to eliminate the frequency drift.
The 1534 nm light comes from a fiber laser and its frequency can be tuned with a piezo.
We up-convert the 1534 nm laser through a periodically-poled-lithium-niobate crystal and then beat the light with the 767 nm laser.
The beat signal is detected and then converted to a feedback signal to lock the piezo of the 1534 nm laser.
The $\pi$ phase shift of the superlattice is calibrated by observing the atom number of the odd or even subsystems after an adiabatic lattice loading, which gives a relative frequency shift of 532.5 MHz.
The beat signal is shown in Fig.\ref{Method:SLphase}(a), where the linewidth at -3 dB below maximum is about 120 kHz, corresponding to a phase broadening of $\delta \varphi = 0.7$ mrad.

The sub-lattice addressing is implemented by introducing a spin-dependent effect onto the superlattice \cite{Yang:2017}.
The short- and long-lattice are set to $84.0(6) E_r$ and $30.8(2) E_r$, respectively.
As the direction of the bias field is along $x$ and the phase of the electro-optical modulator is tuned to $\pi/3$, we create a 28-kHz energy splitting between the hyperfine transition of the odd and even sites.
Then a rapid adiabatical passage is employed to transfer the atoms on odd (even) sites into the hyperfine state $5S_{1/2}\Ket{F=2, m_F=-2}$.
The microwave pulse sweeps with a time-dependent Rabi frequency $\Omega_{\text{MW}}(t)$ and detuning $\delta_{\text{MW}}(t)$,
\begin{equation}
\begin{split}
\Omega_{\text{MW}}(t) = \Omega_0 \text{sech} \left[ \beta \left(\frac{2t}{T_p} -1 \right) \right],\\
\delta_{\text{MW}}(t) = \frac{\sigma_{\text{MW}}}{2} \text{tanh} \left[ \beta \left(\frac{2t}{T_p} -1 \right) \right].\label{eqm1}
\end{split}
\end{equation}
Here $\Omega_0/(2\pi) = 9.85(2)$ kHz is the maximum Rabi frequency, $\sigma_{\text{MW}} = 20$ kHz is the width of the microwave sweep. We set the pulse length to $T_p = 1$ ms and the truncation factor to $\beta =5.3$.
We calibrate the fidelity of spin flips by applying a multi-pulse sequence onto the atomic ensemble.
From the residual density after 10 pulses, we deduce that the fidelity of flipping the odd sites without affecting even sites is 99.5(3)\%.

Using the site addressing technique, we calibrate the zero-point $\varphi = 0$ through the atom tunneling in double-well units \cite{Yang:2017}.
We first prepare a Mott state with near unity filling in short lattices.
Then the atoms in the odd sites are selectively addressed and removed from the superlattice.
We quench the potentials to $16.8(1) E_r$ (short-lattice) and $10.0(1) E_r$ (long-lattice) to allow atom tunneling within double-well units for a half-period (one period is 5.0(1) ms).
Afterwards, we stop the tunneling by ramping up the barrier and detect the atom number of the odd sites.
The tunnelling process is sensitive to the energy bias of the double-wells.

The local superlattice phases are measured via the atom tunneling.
The inhomogeneity of the superlattice phases leads to the irregular stripes at the upper and lower edges of the atomic cloud (Fig. \ref{Fig:f2}A).
We run the sequence at the superlattice depths of $13.7(1) E_r$ (short-lattice) and $10.0(1) E_r$ (long-lattice).
The density distributions are measured after around a half-period of the atom tunneling.
The high density regions in the recorded images Fig. \ref{Method:SLphase}(b) indicate the local degeneracy of the double-well units.
Therefore, we map the superlattice phases into a quadratic-like density pattern, which is smooth in the center and sharp on the edge.
In this sense, the harmonic trap is slightly modified since the energy differential is smaller than the slope of the harmonic envelope in the central region.
At around $30\ \mu$m away from the cloud center, the phase inhomogeneity results in a non-monotonic density distribution.
The harmonic approximation is valid when a smaller cloud is prepared to avoid the outer region.

%%=================================================================================

\subsection{Detection methods}

The number fluctuations on lattice sites are acquired by applying a photoassociation (PA) laser onto the atomic sample.
When the atoms in deep lattices are illuminated by the PA light, two atoms can absorb a photon to form a excited molecular state.
Then the molecule quickly decays to free channels by emitting another photon \cite{Jones:2006}.
The atoms escape from the lattice after this dissociation process since it gives sufficiently high kinetic energy to the particles.
Therefore, only the lattice sites with odd occupations contribute one-atom signal to the final measurements, indicating a parity projection.
Here, we select the PA transition to the $v$=17 vibrational state in the $0_g^-$ channel \cite{Fioretti:2001}, which has a relatively large line strength to drive the intercombination transition $5S_{1/2} \leftrightarrow 5 P_{3/2}$.
The frequency is 13.6 $\text{cm}^{-1}$ red-detuned to the D2 line of $^{87}$Rb atom.

%%=================================================================================

\begin{figure}[!tb]
{\includegraphics[width=8cm]{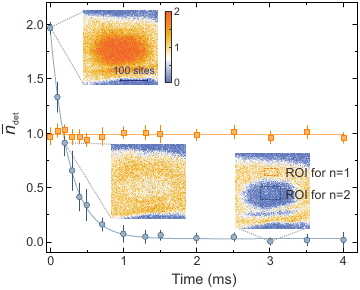}}
  \caption{Photoassociation collision induces loss of atom pairs. The averaged density in the region of interest (ROI) for $n=2$ decays exponentially with a time constant of 180(5) $\mu$s. While the atomic density in the ROI for $n=1$ remains constant. Each data point is a averaged over 20 images and the error bars denote standard deviations.}
\label{Method:PA}
\end{figure}
%%=================================================================================

A home-built diode laser generates the PA light.
This laser is locked onto a Fabry-P\'{e}rot cavity, which is stabilized by the $^{87}$Rb repumper laser.
The PA laser beam has a Gaussian beam waist of 82 $\mu$m at the atomic plane.
We filter the laser with a hot vapor cell of Rb to eliminate the influence of the sidebands at the resonant frequencies, achieving a 30 dB-attenuation.
After atoms are pinned in lattices, we shine the PA light onto the cloud with an intensity of 0.67 W/cm$^{2}$ and then monitor the remaining atoms.
Figure. \ref{Method:PA} shows the decay of a Mott state with initial double filling, giving a decay rate of 5.6(2) kHz.
The one-body scattering rate of the light is 47 Hz.
We can estimate the remaining atoms by taking account of the competition between one- and two-body losses.
The measurements indicate a fidelity of 98.7(3)$\%$ for this parity detection, which agrees well with the theoretical prediction.
Apart from the efficiency of the PA collision, the detection also suffers from the unavoidable shot-noise of the absorption imaging. The duration of the imaging pulse is 10 $\mu$s, giving rise to a standard deviation of 0.6 on the site occupation.
For measuring $\bar{n}$ in the ROI ($10^4$ sites $ = 1780$ pixels), the standard deviation can be suppressed to 0.015.

\begin{figure}[!htb]
{\includegraphics[width=7.1cm]{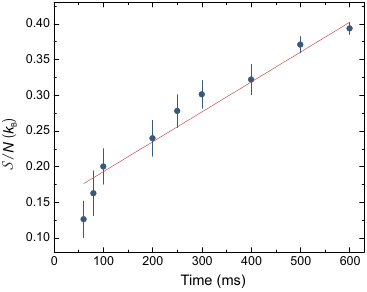}}
  \caption{Heating rate of the optical lattices. The entropy per particle is derived from the remaining densities after the parity projection. The linear fitting has a slope of $0.42(3)\ k_{\text{B}}/s$.}
  \label{Method:heating}
\end{figure}

The entropy of Mott insulator states can be deduced from the parity measurements.
The von Neumann entropy of a quantum system with the density matrix $\rho$ is $\mathcal{S} \equiv - \text{Tr}\{ \rho \log \rho \}$.
In a Mott insulator with $\bar{n} =2$, the lowest-lying excitations are states with filling $n=3$ (particle) and $n=1$ (hole).
We can estimate its entropy using the ``particle-hole approximation'' by neglecting the higher-energy excitations \cite{Gerbier:2007,Ho:2007}.
For the sample without applying the cooling sequence, the temperature of the cloud is mainly controlled by extending the superlattice ramping time.
As shown in Fig. \ref{Method:heating}, the entropy rises as the duration of lattice loading getting longer.
The major heating mechanism is incoherence light scatting imposed on these atoms.
We estimate the heating rate by fitting the data with a linear curve, which results in $0.42(3)\ k_\text{B} /s$ per particle.
%We fit the data with a linear curve and estimate the heating rate to be $0.42(3)\ k_\text{B} /s$ per particle.

%%=================================================================================

%%=================================================================================

%%=================================================================================

\begin{figure*}[!htb]
{\includegraphics[width=12.8cm]{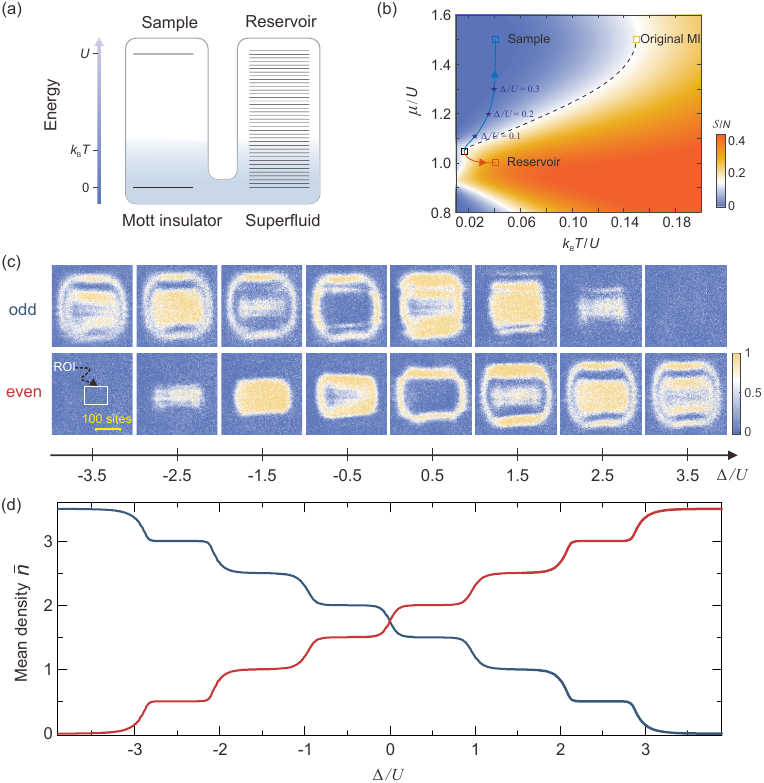}}
  \caption{Immersion cooling. (a) Schematic of the cooling concept. The gapped Mott insulator and gapless superfluid have substantially different density of states. For a low-temperature joint system, the entropy is stored in the reservoir.
  (b) Isentropic route in the entropy diagram. The entropy per particle $\mathcal{S}/N$ of a 1D system at $J/U = 0.02$ is given by the quantum Monte-Carlo simulations.
  The dashed isentropic line links the original $\bar{n}=2$ Mott insulator to a $\bar{n}=1.75$ superfluid state. As $\Delta$ varies from $0$ to $U/2$, the system is divided into a low-entropy Mott insulator and a high-entropy reservoir.
  (c) Realization of the cooling method. We measure the atomic densities of the subsystems at half-integers of $\Delta/U$. In each case, superfluid and Mott insulator phases are coexistent and emerge alternately. We choose a region of interest (ROI) to count the density, which changes from 3.5 (0) to 0 (3.5) in the center of the odd (even) subsystem.
  (d) Original densities before the parity projection. The solid curves are theoretical predictions for the central densities of the odd (blue) and even (red) subsystems at $k_\text{B} T_f/U =0.05$. }\label{Method:Coolingconcept_exp}
\end{figure*}

%%=================================================================================

%%=================================================================================

\begin{figure*}[!htb]
{\includegraphics[width=15cm]{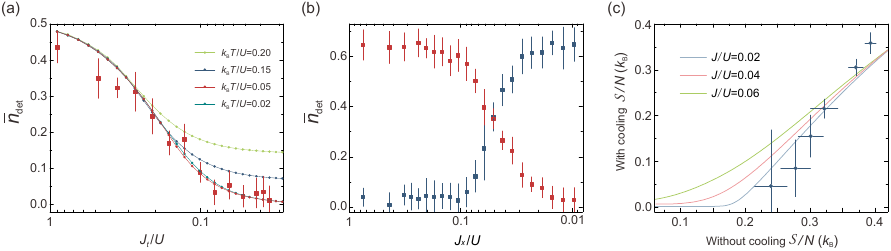}}
  \caption{Phase transitions and critical values. (a) The red points are measurements at different lattice depth during the superfluid to Mott insulator transition. The solid curves are 2D QMC simulations at four different temperatures. (b) The final atomic density is monitored after a sudden change of the energy offset during the lattice loading process. The red and blue data corresponds to the parity projection of the odd and even rows, respectively. The measurements suggest that the partition of atomic density occurs until $J_x/U \sim 0.02$. (c) Cooling power at different thermalization conditions. The solid lines are QMC simulations. The cooling is more efficient when the system can thermalize up to small tunneling strength.}\label{Method:SFMI}
\end{figure*}

%%=================================================================================

%%=================================================================================

\subsection{Quantum Monte-Carlo simulations}

The ultracold Bose gases in optical lattices can be well described by the Bose-Hubbard model \cite{Jaksch:1998}.
Here we use a ``worm'' quantum Monte-Carlo (QMC) algorithm to simulate the equilibrium system \cite{Prokofev1998,Prokofev1998b}.
The simulations are performed on a homogenous 2D square lattices with a size of $10\times 10$, or a 1D chain with a length of $50$.
The system is considered as a grand canonical ensemble.
After thermalization, we continue to perform the QMC sweeps for at least $10^7$ times to get the results under each setting.
When the temperature $T$ and chemical potential $\mu$ are fixed, the atom number fluctuates in the outcome.
For a fixed particle number $\bar{n}$, the chemical potential $\mu$ serves as a variable.

The thermodynamical quantities can be fully determined in this simulation.
From the specific heat $C$, we deduce the entropy density $\mathcal{S}(T)$  as following,
\begin{equation}
\mathcal{S}(T)=\mathcal{S}(T_{_H})-\int_{T}^{T_{_H}} \frac{C}{T'}{\rm d}T'.\label{eqm3}
\end{equation}
To avoid the calculation difficulty at ultra-low temperatures, the integration is carried out from a high temperature $T_{_H}=U$ instead of zero temperature.
The entropy $\mathcal{S}(T_{_H}=U)$ is calculated based on the grand canonical distribution by neglecting the hopping term.
Generally, the simulations are performed in a 2D parameter plane.
We choose the chemical potential $\mu/U$ from 0.8 to 1.6 with a 0.01 interval, and the temperature $k_\text{B} T/U$ from 0.01 to 0.4 with 40 steps.
The entropy per particle at other values on the plane are deduced by performing a 2D interpolation.
The lower boundary of the entropy $\mathcal{S}/N = 1.9^{+1.7}_{-0.4} \times 10^{-3}$ is limited by the accuracy of the simulation at ultra-low temperatures, which also explains the flat part between 0.12 $k_\text{B}$ and 0.15 $k_\text{B}$ in the attached graph of Fig. \ref{Fig:f3}B.

The mean density and number statistics are also obtained.
Comparing to the experimental results at different $\Delta/U$, we fit the measurements based on the particle and entropy conservations.
To acquire the simulation results of Fig. \ref{Fig:f2}B, we perform the calculation of entropy, mean densities and probabilities of odd occupancies in a larger parameter plane.
The chemical potential $\mu/U$ is from -1 to 3.56.
The temperature $k_\text{B} T/U$ starts at 0.01 and end up with 0.3.
At temperature $k_\text{B}T_f/U =0.05$, we find the isentropic cooling trajectory on the entropy and density diagrams.
Vice versa, we can fit the temperature of a cold sample when the mean density and probabilities of the site occupancies are known.

%%=================================================================================

\subsection{Dynamics during the cooling process}

Fig. \ref{Method:Coolingconcept_exp}(a) shows the basic cooling principle, where the superfluid state has much larger density of states than the Mott insulator phase.
In thermal equilibrium, the superfluid phase serves as a reservoir for storing the entropy of the joint system.
We illustrate the entropy redistribution in the entropy diagram of Fig. \ref{Method:Coolingconcept_exp}(b).
The cooling process is illustrated by the isentropic trajectory which links the original $\bar{n} = 2$ Mott insulator and the staggered subsystems with cooling.
Fig. \ref{Method:Coolingconcept_exp}(c) shows the densities at half-integers of $\Delta/U$, where the Mott insulators emerge associated with superfluid states in the other subsystem.

In the cooling samples, we perform a time-resolved measurement on the number statistics across the superfluid-Mott insulator phase transition.
For the averaged filling $\bar{n}=2$, the critical point of the phase transition in a 2D square system is $(J/U)_c =0.035$ \cite{Capogrosso-Sansone:2008,Teichmann:2009}.
The anisotropic tunneling of this staggered system allows us to define a total tunneling strength by neglecting the inter-well coupling  $J_t = J_x+2J_y$.
The strength is $J_t/U = 0.013$ at the end of the lattice loading.
Fig. \ref{Method:SFMI}(a) shows the parity measurements as the tunneling-to-interaction ratio decreases.
Compared to the 2D QMC simulations, the temperature of the samples is $k_B T_f/U \simeq 0.05$.
As the temperature getting even lower, the simulation shows that the value of the observable does not change.
In the reservoir, the superfluid state can survive until the thermal energy becomes larger than the total tunnelling strength $k_B T > J_t$ \cite{Gerbier:2007}.
Since the correlation between the subsystems vanishes gradually as the barrier increases, the entropy of the subsystems has the additive property.

%%=================================================================================

\begin{figure*}[!htb]
{\includegraphics[width=13.6cm]{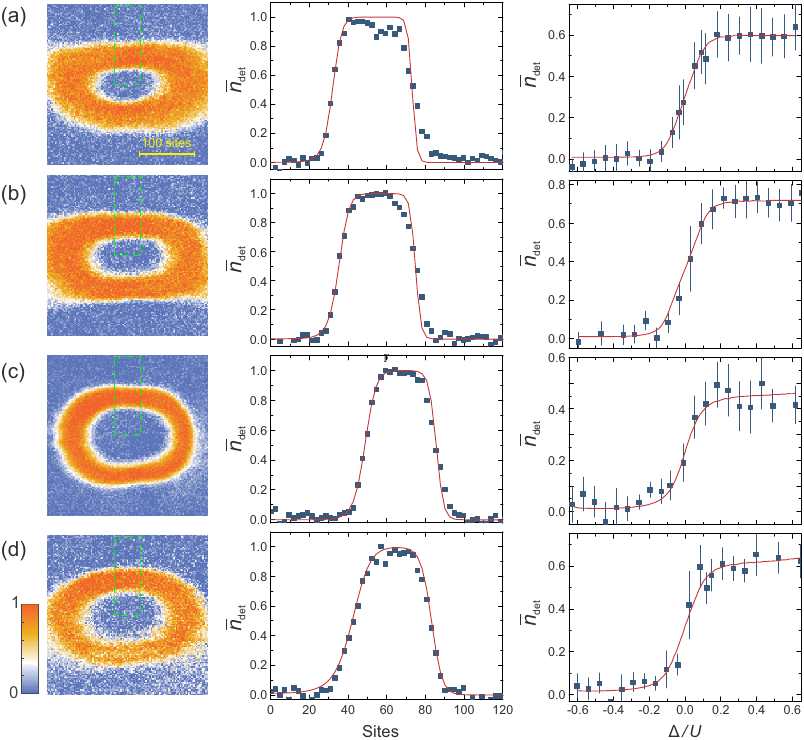}}
  \caption{Thermometry with two methods. (a)-(d) show the cooling samples and thermometers based on two distinct methods. We apply the averaging along $x$ direction to get the density profiles among a region of interest, which are marked by green dashed rectangles. The systems have not achieved a global thermal equilibrium, especially for the low-temperature samples. From the density profiles, we obtain the temperatures of (a)-(d) are $k_B T_f/U = 0.037(4), 0.045(3),0.060(3)$ and 0.083(8), respectively. From the density partition, we get the corresponding temperatures as $k_B T_f/U = 0.038^{+0.007}_{-0.007}, 0.044^{+0.011}_{-0.011},0.073^{+0.008}_{-0.013}$ and $0.087^{+0.007}_{-0.010}$, respectively. These two thermometers are consistent with each other.}
\label{Method:Thermometer}
\end{figure*}

%%=================================================================================

We characterize the thermalization of the system via the quench dynamics.
Generally, the kinetic energy $J$ determines the global thermalization rate of the Hubbard system \cite{Hung:2010}.
While, in the absence of a global mass flow, the equilibration becomes faster in our staggered system.
Here, we reverse the sign of the superlattice phase at a certain time during the loading process to produce a positive energy bias.
The phase switching takes 1 ms, after which the lattice loading continues.
Fig. \ref{Method:SFMI}(b) shows the transport of atoms between the subsystems.
When the change is applied at low lattice depth and the tunneling is larger than $J_x/U=0.1$, the system fully equilibrates and the even rows enter the Mott insulator.
The mass transport becomes slower as the phase change is applied at deeper lattice.
The local mass varies until the tunneling strength becomes $J_x/U=0.02$, where the lattice ramping just has 3 ms left.
In the cooling sequence, we maintain the superlattice phase during the lattice ramping to redistribute the entropy adiabatically.
The $\Delta/U$ in Fig.\ref{Fig:f2}B is calculated under the condition of the final lattice depths $V_s = 26.1(2) E_r$ and $V_l = 10.0(1) E_r$.
The consistence between the experimental results and theoretical predictions (Fig.\ref{Fig:f2}B) confirm that the system can thermalize up to the ending time of the lattice loading.
Here we evaluate the entropy conservation in the staggered system at $J_t/U=0.04$.

Since the cooling power strongly depends on the thermalization rate \cite{McKay:2013}, we present a comparison between the experimental results and theoretical simulations.
For the half-integer filling $\bar{n} =1.5$, the minimum entropy of a classical state is $S_{\text{c}} = k_{\text{B}} \ln2 $.
While this is the upper bound for a superfluid phase due to its underlying quantum correlations.
The value of entropy per particle $\mathcal{S}_{\text{c}}/(2\bar{n})$ defines a critical point for superfluid to normal fluid transition \cite{Trotzky:2010}.
Only when the original entropy becomes lower than this critical value, the onset of the superfluid state in the reservoir leads to strong cooling power.
If the system can only achieve thermal equilibrium at non-zero tunneling strengths, the critical value would becomes smaller \cite{Kantian:2018}.
Fig. \ref{Method:SFMI}(c) shows the cooling effect at different tunneling strengths.
The experimental data are deduced from the parity measurements, which is independent on the newly-developed thermometer.
Compared to the QMC simulations, the measurements support that our system has a fast thermalization and can be characterized as the 1D system at $J/U=0.02$.

%%=================================================================================

\begin{figure}[!htb]
{\includegraphics[width=6cm]{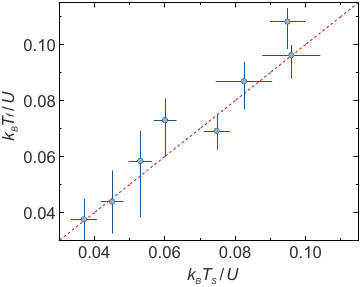}}
  \caption{Comparing the temperatures given by two thermometry methods. The temperatures $T_s$ are acquired by fitting the density distributions, whereas the $T_f$ are obtained via probing the densities at different $\Delta$. The dashed line denotes an equivalence of the temperatures.}
\label{Method_TwoThermometer}
\end{figure}

%%=================================================================================

\subsection{Thermometry}

Here we present three methods for measuring the temperature.
When the many-body wave functions are projected to number states on deep lattices, we can estimate the temperature from the parity measurements.
At high temperatures, such as the entropy without cooling in Fig. \ref{Fig:f3}B, these values are deduced from the probabilities of the odd occupancies.
However, limited by the fidelity of the parity detection, we probe the low-temperature system with two other approaches.
One is based on the density partition and the other relies on the shell structure of the Mott insulator state \cite{Bakr:2010,Sherson:2010}.
Fig. \ref{Method_TwoThermometer} shows the consistency of these two methods, and the experimental details are presented below.

By further evaporating the 2D gas,we carry out these two thermometries on a small cloud with $\sim 4 \times 10^4$ atoms.
The transverse confinement is provided by a Gaussian potential.
With the local density approximation, the chemical potentials of the lattice sites are $ \mu \left( y \right) = \mu \left( 0 \right) - V \left( y \right)$, where $\mu(0)$ is the chemical potential at the trap center.
Fig. \ref{Method:Thermometer} shows the thermometry of these two methods performed on the cooling samples.
The temperature of the system is controlled via the duration and adiabaticity of the lattice ramping.
We find that the system has lower temperatures when the lattice loading has smaller modifications on the cloud size.
This is because the adiabaticity is improved as the particles undergo less transport \cite{Dolfi:2015}.
The whole cloud does not reach a global thermal equilibrium.
The thermometry based on the partition of density is robust to the irregular trap potential and can be used to probe systems without a global thermal equilibrium.
Furthermore, via counting the number statistics of the Mott insulator state, we can apply this thermometry onto a superfluid phase by coupling it with an insulating state.

%%=================================================================================

\begin{figure*}[!htb]
{\includegraphics[width=12cm]{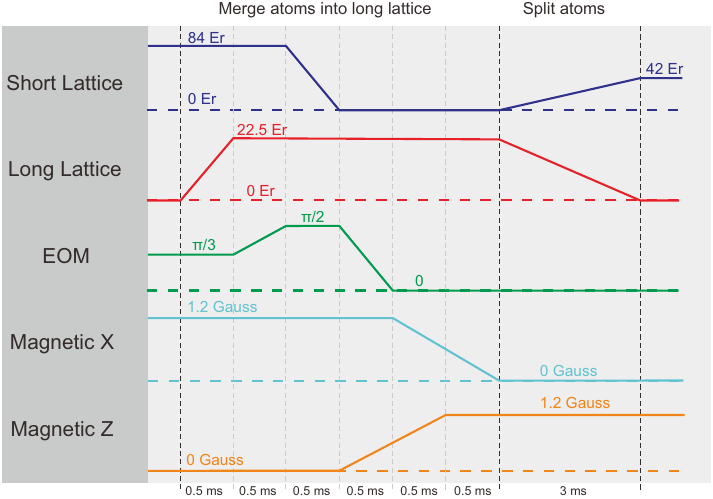}}
  \caption{Experimental sequence for atom redistribution in the double-well units. After the superlattice potential are ramped up, we adjust the EOM to $\pi/2$ to shift the atom pairs to the potential minimum of the long-lattice. Then the short-lattice is reduced to map the atom pairs into the ground state of the long-lattice. Afterwards, the axis of the magnetic field is changed from $x$ to $z$ direction. Making use of the on-site interaction, we split the atoms into double wells by ramping the lattices in 3 ms.}
\label{Method:Seq}
\end{figure*}

%%=================================================================================

\subsection{Quantum state engineering in superlattice}

In the insulating region, the defects on lattice sites could be reduced by engineering the Hubbard interactions within few-body systems \cite{Rabl:2003,Popp:2006,Cheinet:2008,Bakr:2011,Tichy:2012}.
Here, we redistribute the atom pairs into single occupancy within double-well units.
After the cooling sequence, the particles in the superfluid reservoir are removed by applying site-selective addressing sequences, leaving behind a filling-2 Mott state at odd rows.
Then we perform an operation to arrange the $|2,0\rangle$ to $|1,1\rangle$, as shown in Fig. \ref{Method:Seq}.
The atoms are first transferred from the odd sites into long-lattice potential, secondly the atoms are split into single occupancy.
In the first step, tuning the short-lattice potential to match the wave functions of the long-lattice can avoid the atom excitations to higher bands.
In the second step, the inhomogeneity of the superlattice phase leads to a competition between the interaction blockade and the energy bias at the regions away from the cloud center, which explains the appearance of fringes in the attached graphes of Fig. \ref{Fig:f3}C.
In addition, the intrinsic heating of the lattice laser plays a role as the operation time becomes longer.
Finally, we optimize the operation by suppressing these excitations.

We calibrate the fidelity of the operation with parity detection method.
The parity projection remove the atom pair of $|2,0\rangle$, while it has no impact on the $|1,1\rangle$ state.
Since $|2,0\rangle$ has negligible decay during the operations, the error of the operation mainly comes from the crosstalk between different double-well units.
This error accumulates and its fidelity $\eta$ can be evaluated from the final states after applying multi-cycles $N_1$.
The decay of the state $|1,1\rangle$ follows a power law function $\eta^{N_1}$ and the $|2,0\rangle$ changes in the form of $1-\eta^{N_1}$.
We fit the measurements with these functions and obtain a fidelity of $\eta = 99.3(1) \%$.
The final fidelity of the unity filling among $10^4$ sites is $99.2(1) \%$ by taking account of the site occupation of the cooling samples.

\subsection{High-fidelity two-qubit quantum gate}

Entangling neutral atoms have been achieved with several types of interactions.
The contact on-site interaction between atoms induces a collision phase \cite{Jaksch:1999}, which has been used for creating entanglement in state-dependent optical lattices \cite{Mandel:2003,Anderlini:2007}.
Superexchange interaction can entangle the spin states of neighboring atoms in a controllable way towards large scales \cite{Duan:2003,Trotzky:2008,Vaucher:2008,Dai:2016}.
Besides that, long-range Rydberg interactions can significantly reduce the gate operation time \cite{Jaksch:2000,Isenhower:2010,Levine:2018}, which has enabled the generation of a 20-qubit ``Schr\"{o}dinger cat'' state recently \cite{Omran:2019}.
However, the fidelity of the logical gates with neutral atoms has not exceed the threshold required by fault-tolerant schemes.
Generally, the gate fidelity depends on the relative time scale between the gate operation and the state coherence.
Here, we enhance the interaction strength of superexchange to half of the on-site interaction by engineering the Hubbard Hamiltonian in isolated double-well units.
Moreover, the coherence time of the gate operation is extended to approaching the limit of the light scattering.
Therefore, we achieve a two-qubit gate with error rate below the error threshold of fault tolerance and generate massive pairs of Bell states in the double-well superlattice.

As shown in Fig.\ref{Fig:f1}C, we consider two bosonic atoms with different hyperfine states denoted by pseudo spins $\ket{\uparrow}$ and $\ket{\downarrow}$ in a symmetric doule-well (DW) unit.
The state of this system can be decomposed into four Fock states as $\{ \ket{\uparrow, \downarrow}, \ket{\downarrow, \uparrow}, \ket{\uparrow\downarrow,0}, \ket{0,\uparrow\downarrow} \}$.
When the initial state has one atom per site (such as $\ket{\uparrow, \downarrow}$ or $\ket{\downarrow, \uparrow}$), the anti-symmetric eigenstate $(\ket{\uparrow\downarrow,0} - \ket{0,\uparrow\downarrow})/\sqrt{2}$ is decoupled from the other eigenstates in the Hubbard Hamiltonian \cite{Trotzky:2008}.
While the symmetric eigenstate $\ket{D}=(\ket{\uparrow\downarrow,0} + \ket{0,\uparrow\downarrow})/\sqrt{2}$ need to be considered.
The Hamiltonian can be written as,

\begin{equation}
\hat{H} = \begin{array}{@{}r@{}c@{}c@{}c@{}c@{}l@{}}
    & \Ket{s} & \Ket{t} & \Ket{D}  \\
    \left.\begin{array}
    {c} \Ket{s} \\\Ket{t} \\\Ket{D}
    \end{array}\right(
                    & \begin{array}{c} 0 \\ -\Delta_s \\ 0  \end{array}
                    & \begin{array}{c} -\Delta_s \\ 0 \\ -2J \end{array}
                         & \begin{array}{c} 0 \\ -2J \\ U \end{array}
                          & \left)\begin{array}{c} \\ \\ \\ \end{array}\right.
  \end{array}
\end{equation}

Here, the spin-dependent energy between the sites of a DW is represented by $\Delta_s$, which is tuned to 0 during the spin dynamics.
The Hilbert space contains three eigenstates, where the coefficients in the eigenstates are $\alpha=4J/(U + \sqrt{16 J^2 + U^2})$, $\beta = 4J/(U- \sqrt{16 J^2 + U^2})$.
The spin dynamics in DWs can be predicted by the Schr\"{o}dinger equation.
In the regime of $J \ll U$, the probability of the intermediate state $\ket{D}$ with double occupancy on a lattice site is negligible small.
Hence the spin dynamics is governed by the superexchange process with a weak interaction strength of $4J^2/U$.
If we enter a regime where $J \sim U$, the superexchange dynamics becomes much faster and the population on the state $\ket{D}$ cannot be neglected.
Additionally, the system is described by the generalized Bose-Hubbard model, where some effects like density-induced tunneling should be considered \cite{Trotzky:2008}.
However, when the Hubbard parameters are set to $J/U=\sqrt{3}/4$, the frequencies satisfy a relation of $\omega_2 = 4 \omega_1$.
After a time interval of $\pi/(2\omega_1)$, the initial state $\ket{\uparrow, \downarrow}$ evolves to an entangled state $(\ket{\uparrow, \downarrow} + i \ket{\downarrow, \uparrow})/\sqrt{2}$.
By tuning the lattice potentials precisely, we realize this two-qubit $\sqrt{\text{SWAP}}$ gate within thousands double wells.

%%=================================================================================

\begin{figure*}[!htb]
{\includegraphics[width=14cm]{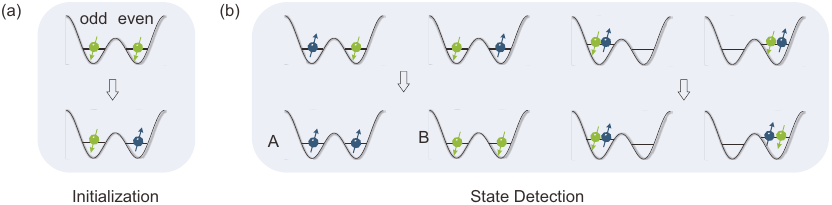}}
  \caption{Experimental sequence for state initialization and detection. (a) For preparing the initial state, we flip the all the spins of the even sites from $\ket{\downarrow}$ to $\ket{\uparrow}$. (b) After the spin dynamics, we detect the spin states by addressing all the even sites and then measuring the spin components with $in\ situ$ absorption imaging (A and B). For the spin state $\ket{\uparrow,\downarrow}$ ($\ket{\downarrow,\uparrow}$), the signal is recorded by the image A (B). While the states $\ket{\uparrow\downarrow,0}$ and $\ket{0,\uparrow\downarrow}$ contribute equally to the densities on images A and B. }
\label{Method_Initialization_Detection}
\end{figure*}

%%=================================================================================

The state initialization and detection are illustrated in Fig.\ref{Method_Initialization_Detection}.
Controlling the depths of the bichromatic lattices allows one to tune the DW structure in terms of the Hubbard parameters $J$ and $U$.
Starting from the states $\ket{\downarrow,\uparrow}$, the spin dynamics is initialized by linearly ramping down the short-lattice from 50.0(3) $E_r$ to 9.76(1) $E_r$.
The long-lattice is kept at 10.0(1) $E_r$ to block the coupling between distinct DWs.
Within each DWs, the tunneling rate is $J \simeq$ 613 Hz and the on-site interaction is $U$ = 1.41(1) kHz.
Fig. \ref{Fig:f4}C shows the time evolution of the state populations on image A.
Atoms in DWs are entangled and disentangled consecutively.
The fitting result gives a ratio of the oscillation frequencies as $\omega_2/\omega_1$ = 3.99(1).
The time interval for performing the gate operation becomes 71 times shorter than the previous result \cite{Dai:2016}, which explains the significant improvement of the gate fidelity.
However, the oscillation shows a shorter coherence time 57(16) ms which is partially caused by the spatial inhomogeneity of the interaction strength among the sample \cite{Dai:2016}.
Considering the Gaussian profile of the lattice beams, the averaging over 1250 DWs contributes to a decay time of 120 ms.
In addition, fluctuation of the superlattice phase also leads to part of the decoherence.

Besides the spatial imhomogeneity and noises, the photon scattering induced by lattice lasers limits the coherent time of the entangled states.
The scattering rate of the atomic internal states is proportional to the lattice depth $V$, such as $\Gamma_0 = \Gamma V/\hbar \Delta_L$ \cite{Grimm:2000,Pichler:2010}.
Here $\Gamma$ is the natural line width of the transition and $\Delta_L$ is the laser detuning.
However, in optical lattice, the discrete energy bands generally reduce the coupling between different motional states, which is characterized by the Lamb-Dicke parameter $\eta_{LD}$.
We find only the light scattering that changes the motional quantum states would destroy the entanglement.
For the atoms occupied on the ground band, the total scattering rate to the other excited bands can be deduced as $\Gamma_{\mathrm{t}}\!=\!\Gamma_0(1- e^{-\eta_{LD}^2})$ \cite{Wineland:1998}.
The probability to excite atoms to other bands becomes strongly suppressed when we enter the deep Lamb-Dicke regime with $\eta_{LD} \ll 1$.
In this model, $N$-particle entanglement would have shorter lifetime scaled as $1/N$ due to the light scattering.

%%=================================================================================

\begin{figure}[!htb]
{\includegraphics[width=7.3cm]{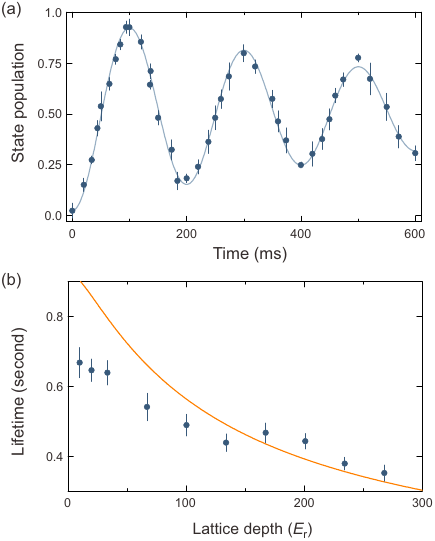}}
  \caption{State coherence and light scattering. (a) A low-frequency superexchange dynamics. Each data are averaged over 5 experimental realizations. A damped sinusoidal fitting indicates a oscillation frequency of $4.99(3)$ Hz and a lifetime of 646(32) ms. (b) Measurement of the lifetime at different depth of the pancake lattice. The solid curve is a theoretical calculation of the system by considering the decoherence due to light scattering. }
\label{Method_Coherence}
\end{figure}

%%=================================================================================

Generally, the atoms can be trapped in optical lattice with a lifetime of $\sim$10 s.
The light scattering mostly destroy the spin correlations of entangled states.
Therefore, we study the decoherence mechanism by measuring the spin-exchange dynamics under different depths of the $z$-lattice.
The superexchange interaction is tuned to be several Hertz to avoid the influence of the spatial inhomogeneity.
We set the depths of the short- and long-lattice to 17.3(2) $E_r$ and 10.4(1) $E_r$, respectively.
Fig. \ref{Method_Coherence}(a) shows a exemplary spin dynamics under the Hubbard parameters of $J \simeq 108(1)$ Hz, $U=1.69(1)$ kHz.
The lifetime of the spin dynamics is extracted by fitting the measurements with a damped sinusoidal function.
In Fig. \ref{Method_Coherence}(b), the coherence time decreases as the lattice potential increases because of the enhancement of the scattering rate $\Gamma_{\mathrm{t}}$.
The measurements are in consistence with the theoretical predictions in deep lattices.
In the shallow lattices, the long-range magnetic dipole-dipole interaction \cite{Lahaye:2009} leads to a further reduction of the state coherence.
By reducing the phase and intensity noise of the lattice lasers, we extend the lifetime by a factor of 2.3 comparing with the previous results \cite{Dai:2016}.
Moreover, this lifetime is 1700 times longer than the duration of the two-qubit gate operation, leaving space for further improvement of the gate fidelity.

The fidelity of the experimentally produced density matrix $\rho_{\text{exp}}$ can be written as,
\begin{equation}
\begin{split}
F &=  \bra{t}\rho^{\text{exp}}\ket{t} \\
&= \frac{1}{2} (\rho^{\text{exp}}_{d,d} + \rho^{\text{exp}}_{b,b}+ \rho^{\text{exp}}_{d,b}+\rho^{\text{exp}}_{b,d} ),
\end{split}
\label{Eq4}
\end{equation}
where the brightness of the image A is used to define the Fock states as $\ket{d}=\ket{\uparrow,\downarrow}$, $\ket{b}=\ket{\downarrow,\uparrow}$.
The diagonal components $\{\rho^{\text{exp}}_{d,d}, \rho^{\text{exp}}_{b,b}\}$ of the density matrix can be obtained by measuring the probability of the two-qubit spin state $P_d$, $P_b$.
The off-diagonal elements $\{\rho^{\text{exp}}_{d,b}, \rho^{\text{exp}}_{b,d}\}$ can be determined by the spin-correlation measurement via the ${\sigma_x,\sigma_y}$ spin rotations.
Then the sum of the off-diagonal terms in Eq.\ref{Eq4} is $\rho^{\text{exp}}_{d,b} + \rho^{\text{exp}}_{b,d} = \frac{1}{2}\left(\langle\sigma_x \sigma_x \rangle + \langle \sigma_y \sigma_y \rangle\right)$.
Since the spin-flip error is negligible small, the error of the two-qubit gate mainly comes from the phase decoherence and the double occupied states $\ket{\uparrow\downarrow,0}, \ket{0,\uparrow\downarrow}$.

We start the gate operator by linear ramping down the short-lattice in $t_1$ to enable the spin dynamics, as shown in Fig. \ref{Fig:f4}C.
After holding the atoms in the DWs for a time duration of $t_2$, the dynamics is stopped by linear ramping back the short-lattice potential.
Such a complete gate operator has the same effective interacting time as a quarter period of the superexchange process in Fig. \ref{Fig:f4}B.
The linear ramp of the lattice potential in the gate operation cause negligible error to the quantum states.
The achieved fidelity is $99.3(1) \%$ from the fitting results.
Afterwards, we turn on a staggered magnetic gradient to rotate the phase of the state $(\ket{d} + i \ket{b})/\sqrt{2}$ to the target Bell state $\ket{t}=(\ket{d} + \ket{b})/\sqrt{2}$ \cite{Dai:2016,Yang:2017}.

The spin correlation is measured on the bases of $\ket{\uparrow/\downarrow}$, $\ket{+/-}$ and $\ket{\circlearrowright/\circlearrowleft}$, respectively.
By adiabatically removing the barrier inside the DWs, these two atoms enter the same long-lattice sites.
Then we shine a photoassociation laser onto the sample to remove the atom pairs on the spin state of $\ket{\downarrow\downarrow}$, with which the probability of the state is measured by counting the atom loss \cite{Dai:2016}.
The other spin components are measured by transferring them into the $\ket{\downarrow\downarrow}$ state.
After 29 consecutive gate operations, the expectation values of the spin correlations are $\langle\sigma_x \sigma_x \rangle = P_{\ket{+,+}} + P_{\ket{-,-}} - P_{\ket{+,-}} - P_{\ket{-,+}}$ = 0.75(3) and $\langle\sigma_y \sigma_y \rangle = P_{\ket{\circlearrowright,\circlearrowright}} + P_{\ket{\circlearrowleft,\circlearrowleft}} - P_{\ket{\circlearrowright,\circlearrowleft}} - P_{\ket{\circlearrowleft,\circlearrowright}}$ = 0.75(4).
The fidelity of this state after 29 gates is $ F^{29} = \frac{1}{2}(P_d + P_b) + \frac{1}{4}(\langle\sigma_x \sigma_x \rangle + \langle\sigma_y \sigma_y \rangle) = 0.80(2)$.
For those atom pairs correctly prepared in the $\ket{\downarrow,\uparrow}$ state, they are entangled with $F = 99.3(1)\%$ fidelity.

\begin{table}[!htbp]
\centering
{\includegraphics[width=8cm]{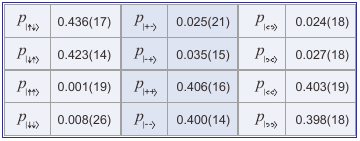}}
\caption{Probabilities of the spin correaltions.}
\end{table}

A full quantum state tomography would be advantageous for evaluating the performance of the entangling gate.
The light scattering \cite{Grimm:2000,Pichler:2010} and magnetic dipolar interaction \cite{Lahaye:2009} represent the fundamental limits for the coherence of the neutral atoms in optical lattice.
These could be improved by using far-detuned lasers and magnetic insensitive states.
Nevertheless, the coherence time constrained by these limitations is still 10$^3$ times longer than the gate operation.
Thus the fidelity of the two-qubit gate can be improved by reducing the spatial inhomogeneity and the laser noise.
In addition, one could increase the interaction strength and thereby reduce the gate operation time by increasing the on-site interaction via magnetic Feshbach resonance \cite{Chin:2010}.
Further developments on quantum state detection \cite{Bakr:2009,Sherson:2010} and high-fidelity single-qubit gate \cite{Weitenberg:2011,Wang:2016} will be needed for quantum information processing.

%%=================================================================================

%%=================================================================================

\end{document}